\documentclass[12pt]{article}
\usepackage{amsfonts}
\usepackage{amsmath, graphicx, epsfig,psfrag, verbatim}
\usepackage{eucal,  amssymb,amsmath,amsthm,graphicx, amsfonts, latexsym,xcolor}
\usepackage[utf8]{inputenc}
\newcommand\btau{\boldsymbol{\tau}}

\begin{document} \parskip=5pt plus1pt minus1pt \parindent=0pt
\title{The risk for a new COVID-19 wave -- and how it depends on $R_0$, the current immunity level and current restrictions}
\author
{Tom Britton$^{1\ast}$,  Pieter Trapman$^{1}$ and Frank Ball$^{2}$ \\
\\
\normalsize{$^{1}$Department of Mathematics, Stockholm University,  Sweden}\\
\normalsize{$^{2}$School of Mathematical Sciences, University of Nottingham,  UK.}\\
\\
\normalsize{$^\ast$To whom correspondence should be addressed; E-mail: tom.britton@math.su.se}
}
\date{\today}
\maketitle

\begin{abstract}
The COVID-19 pandemic has hit different parts of the world differently: some regions are still in the rise of the first wave, other regions are now facing a decline after a first wave, and yet other regions have started to see a second wave. The current immunity level $\hat i$ in a region is closely related to the cumulative fraction infected, which primarily depends on two factors: a) the initial potential for COVID-19 in the region (often quantified by the basic reproduction number $R_0$), and b) the timing, amount and effectiveness of preventive measures put in place. By means of a mathematical model including heterogeneities owing to age, social activity and susceptibility, and allowing for time-varying preventive measures, the risk for a new epidemic wave and its doubling time, and how they depend on $R_0$, $\hat i$ and the overall effect of the current preventive measures, are investigated. Focus lies on quantifying the minimal overall effect of preventive measures $p_{Min}$ needed to prevent a future outbreak. The first result shows that the current immunity level $\hat i$ plays a more influential roll than when immunity is obtained from vaccination. Secondly, by comparing regions with different $R_0$ and $\hat i$ it is shown that regions with lower $R_0$ and low $\hat i$ may now need higher preventive measures ($p_{Min}$)  compared with other regions having higher $R_0$ but also higher $\hat i$, even when such immunity levels are far from herd immunity.
 
\end{abstract}

\footnotetext[1]{Stockholm University, Department of Mathematics, Sweden. E-mail: tom.britton@math.su.se}

\section*{Introduction}\label{sec-Intro}

COVID-19 is currently spreading in many parts of the world. In several regions the spreading has now dropped substantially, and in other regions the first wave has been very small, most often owing to the implementation of  effective preventive measures. Regions where spreading now is low face two competing interests: lifting restrictions to normalize society, and to keep or strengthen restrictions in order to avoid a new major COVID-19 outbreak. The minimal overall effect of preventive measures needed to avoid a large future outbreak, denoted $p_{Min}$, and how it depends on the basic reproduction number $R_0$ and the current immunity level $\hat i$, is hence a highly important question which is investigated here. In addition we consider the doubling time of a new outbreak should it take place, which gives an indication of its impact before additional preventive measures would be implemented. 

The basic reproduction number $R_0$ quantifies the initial potential of an epidemic outbreak for a particular disease in a particular region, and is defined as the \emph{average} number of new infections caused by a \emph{typical} infected individual in the beginning of the epidemic outbreak (before preventive measures are put in place and before population immunity starts to build up), \cite{DHB13}. For COVID-19 estimates of $R_0$ vary substantially between different regions, e.g.\ between 2 and 5 among 11 European countries \cite{FMG20publ}.

Preventive measures aim to reduce the average number of infections caused by an infective, by either reducing the risk of transmission given a contact (e.g.\ hand washing, wearing face mask), reducing the number of daily contacts (e.g.\ social distancing, school closure) and/or reducing the effective infectious period (e.g.\ testing and isolating, treatment). Let $p(t)$ denote the overall effect of such preventive measures at time $t$, where $0\le p(t)\le 1$, and  with $p(t)=0$ corresponding to no preventive measures and $p(t)\approx 1$ meaning more or less complete isolation of all individuals.

Let $\hat i(t)$ denote the community fraction that cannot get infected at time $t$, a few of these being currently infectious, but the majority having recovered from the disease and now being immune (waning of immunity is here neglected since our time frame is less than a year). At time $t$, it is the current (or effective) reproduction number $R_t$ of a region that determines if a new main outbreak can take place or not. In particular, a region with low current transmission avoids the risk for a large new outbreak as long as $R_t< 1$, and regions with ongoing transmission will see a decline in transmission whenever $R_t <1$.

For simple epidemic models, which assume a homogeneous community that mixes homogeneously, it is well-known that $R_t=R_0(1-p(t))(1-\hat i(t))$, since $R_0$ is reduced both due to the preventive measures and from the fact that some contacts will be with already infected people. This implies that $R_t \le 1$ is equivalent to  $p(t)\ge 1-1/(R_0(1-\hat i(t)))$, thus quantifying, in terms of $R_0$ and the current immunity level $\hat i(t)$, the minimal amount of preventive measures needed to avoid a new large outbreak.

For more realistic epidemic models this simple relation between $R_t$ and $R_0$, $p(t)$ and $\hat i(t)$ does not hold. In fact, for epidemic models acknowledging population heterogeneities it holds that $R_t < R_0(1-p(t))(1-\hat i(t))$. The main reason is that individuals having high social activity and/or high susceptibility are more likely to be infected early in the epidemic, implying that individuals at risk later in the epidemic will on average be less susceptible and socially active, thus also infecting fewer if they become infected \cite{BBT}. Here we study an epidemic model in which social mixing depends on age-structure, that also allows for variable social activity as well as variable susceptibility within age-groups. For this model the aim is to quantify $R_t$ as a function of $R_0$, $p(t)$ and $\hat i(t)$, and in particular to quantify the minimal amount of restrictions $p_{Min}$ for a region having initial basic reproduction number $R_0$ and current immunity level $\hat i$ (the index $t$ is now dropped and implicitly considered as current time).

We illustrate our findings by expressing $p_{Min}$ and the doubling time for different regions in Europe and US, but these illustrations are by no means exact. First, the model is clearly a simplification of the ongoing COVID-19 pandemic, but even more so the estimates of $R_0$ and the current immunity level $\hat i$ for different regions contain appreciable uncertainty. Nevertheless our results allow, for the first time to our knowledge, a risk comparison between regions having  different $R_0$ and different immunity level $\hat i$.

\section*{An epidemic model with age-cohorts, variable social activity and variable susceptibility}

The epidemic model is based on the model in Britton et al.\ \cite{BBT}. Individuals are divided into 6 different age groups, and mixing patters are taken from the empirical study of Wallinga et al.\ \cite{WTK06}. Within each age group individuals are divided into three categories: 50\% have normal social activity, 25\% have low social activity (half as many contacts as those with normal activity) and 25\% have high social activity (double activity). It is important to stress that social activity affects both the risk of getting infected \emph{and} infecting others in that socially active individuals have more contacts both when susceptible and when being infectious.

To this model with age-cohorts and variable social activity, studied in \cite{BBT}, we now also add variable susceptibility \cite{G20}, which is done similarly to variable social activity. We assume that 50\% have normal susceptibility, 25\% have half the susceptibility and 25\% are twice as susceptible, and the variable susceptibility is assumed to be independent of both social activity level and age-group. The choice to divide social activity as well as susceptibility into three groups as above is of course quite arbitrary. This choice of heterogeneity structure is quite moderate in that there is no tail (with individuals having very high social activity or susceptibility) and the coefficient of variation equals 0.48 which is moderate (see the Supplementary Materials, SM, for further comments).

It seems natural to also add variable infectivity for individuals who become infected. However, such variable infectivity will have no effect on our results if it is assumed independent of susceptibility and social activity, and is hence omitted.

We use a deterministic SEIR epidemic model (see SM) with a total
of 6*3*3=54 different types of individuals, but very similar results would be obtained from simulations of a corresponding stochastic model assuming a large population (which can be proved using methods in \cite{Ethi09}). The latent state ''E'' (for exposed) is assumed to have mean 3 days followed by an infectious period (''I'') having mean 4 days, thus being quite close to other models for the spread of COVID-19 \cite{FMG20publ}. Details of the model are given in the SM, where it is explained that our results hold also for the corresponding model in which the latent and infectious periods need not follow exponential distributions, and more generally for the model in which infectives have independent and identically distributed shapes of the infectivity profiles. 

It is straightforward to numerically derive properties of the model, such as the basic reproduction number $R_0$, the time dynamics and its final fraction infected when the epidemic stops (see SM for further details).

\section*{Prevention}

During the outbreak, preventive measures of varying magnitude may be put in place. We assume that these preventive measures do not affect the latent and infectious periods, but only that they reduces the rate of infectious contacts. More precisely we make the strong and somewhat restrictive assumption that, at time $t$, all contacts (between the 54 types of individuals) are reduced by the \emph{same} factor $p(t)$. This assumption can easily be relaxed, but to explore all possibilities of contact reduction is infeasible, and among all specific preventions the uniform one, where all contacts are reduced by the same factor, is the most natural choice. Even when assuming such uniform reduction of contact rates, its reduction may vary in time in different ways. However, in the SM it is shown that the exact time allocation of the preventive measures has negligible effect in our model: any time-varying preventive measures $\{p(t); 0\le t \le t_0\}$ from the start of the epidemic up until some fixed time $t_0$, leading to the same overall fraction infected, will have nearly the same fractions infected among the different types of individuals (see SM for details). Thus early mild preventive measures will result in the same composition of infected individuals as doing nothing and then suddenly going to a full lockdown, assuming the two preventive measures lead to the same overall fraction infected.

\section*{The minimal amount of preventive measures $p_{Min}$}

Consider a large community in which COVID-19 spreads according to our model with some fixed value of $R_0$, and for which preventive measures $\{p(t); 0\le t \le t_0\}$ were put in place (the same preventive effect on all type of contacts). We further assume that by $t_0$ the transmission has more or less stopped, resulting in a fraction $\hat i$ having been infected (and are immune) and the remaining fraction $1-\hat i$ still susceptible. Our main scientific question lies in quantifying what the effective reproduction number $R_{t_0}$ equals if all restrictions are lifted at time $t_0$. If $R_{t_0}>1$ it follows that a new large epidemic outbreak may occur if all restrictions are lifted, as opposed to the case $R_{t_0}\le 1$ when herd-immunity has been reached (though smaller local outbreaks are still possible).

\begin{figure}[ht]
\begin{center}
\includegraphics[scale=0.5, angle=0]{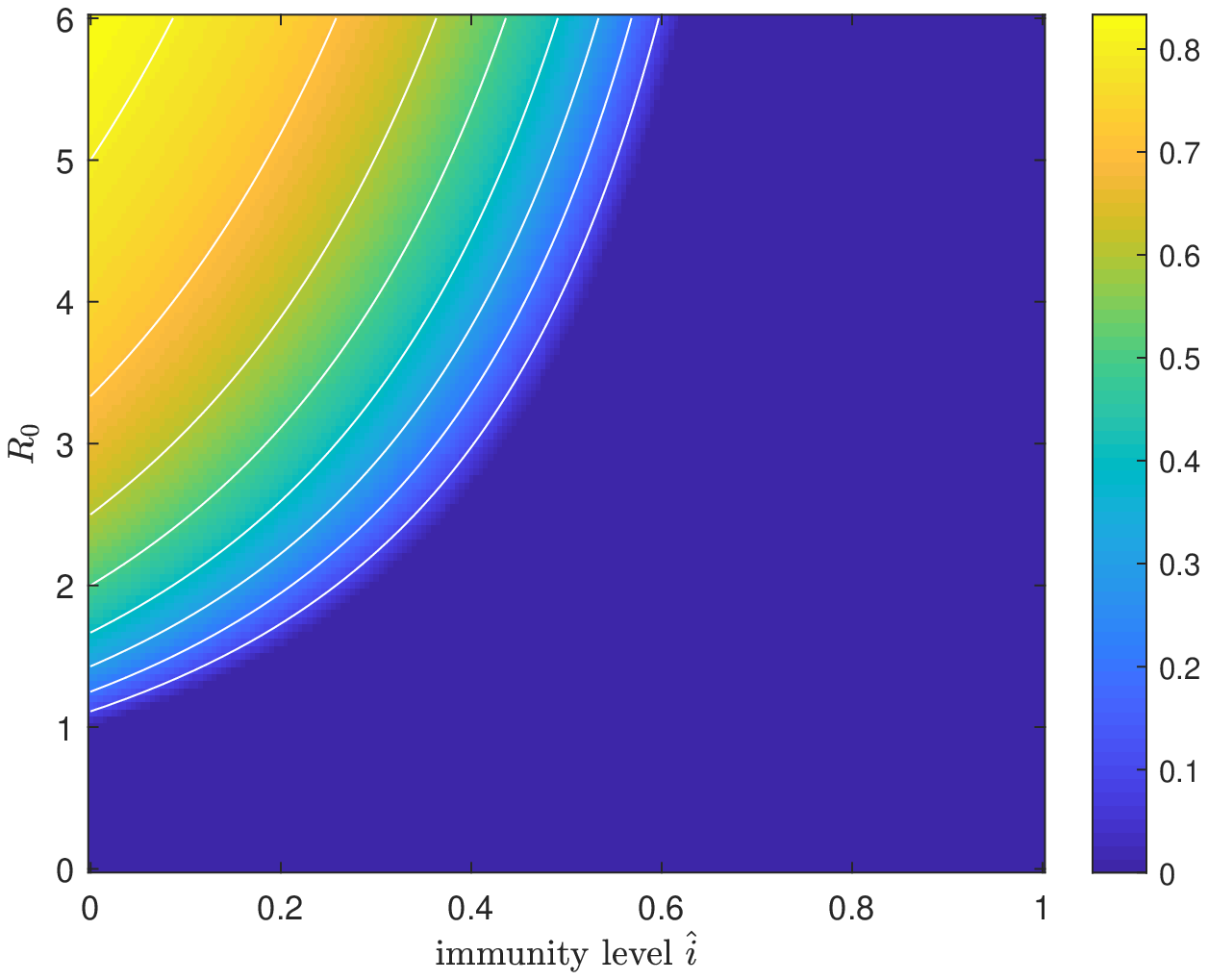}
\includegraphics[scale=0.5, angle=0]{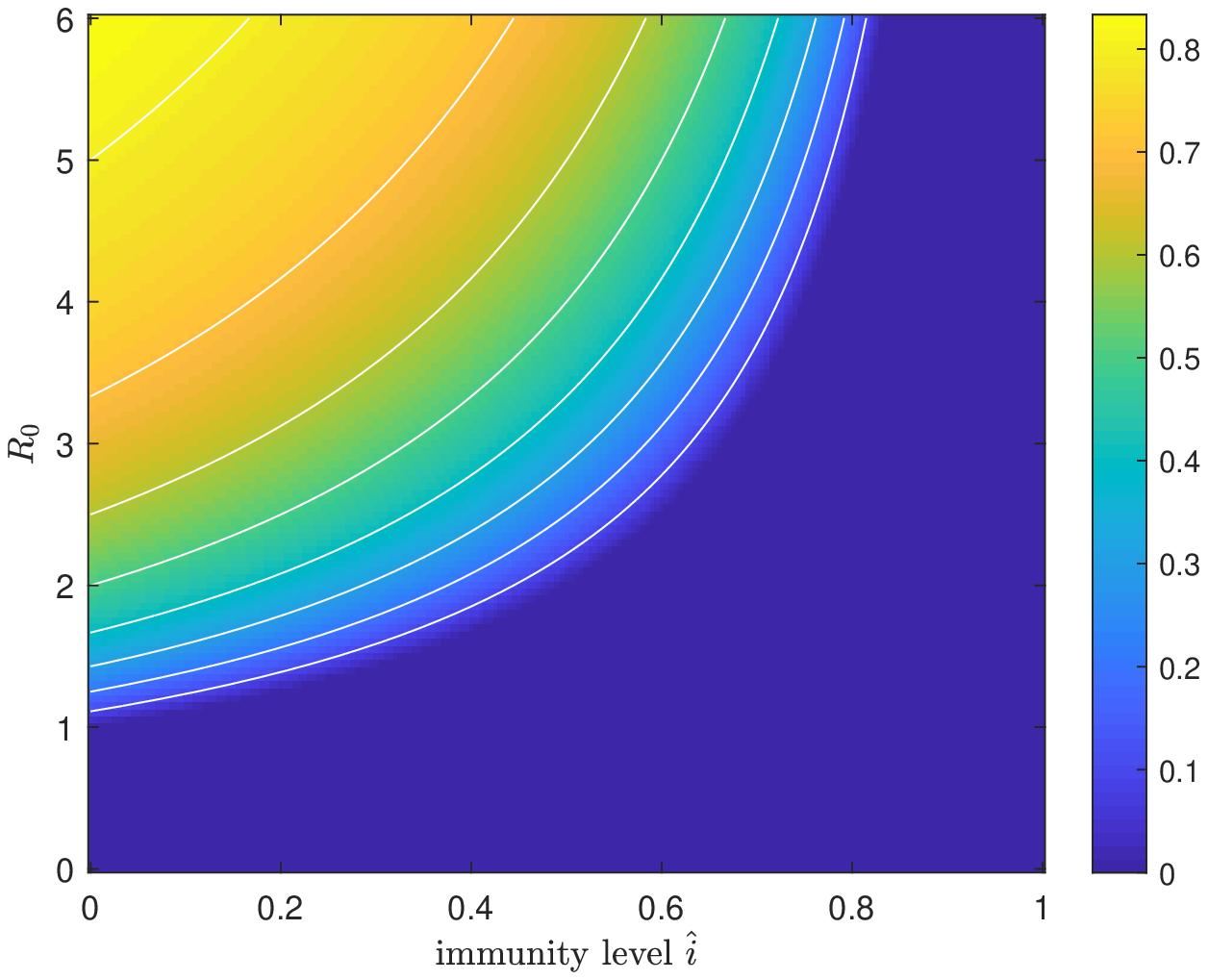}
\end{center}
\caption{Plot of the minimal amount of preventive measures, $p_{Min}$, necessary to avoid a new large outbreak, as a function of $R_0$ and the current immunity level $\hat i$. The left plot is for disease-induced immunity and the right plot is for vaccine-induced immunity.}
\label{p_Min_full-mod}
\end{figure}

In the more common COVID-19 scenario that $R_{t_0}>1$, the minimal amount of preventive measures necessary to avoid a new large outbreak is given by $p_{Min}=1-1/R_{t_0}$. This amount $p_{Min}$ is thus a measure for the risk of a new large outbreak. In the plots below $p_{Min}$ has been computed as a function of $R_0$ and the current immunity level $\hat i$, and is quantified by a heatmap. The left in Figure \ref{p_Min_full-mod} is for the main model allowing for heterogeneities with respect to age, social activity and susceptibility, and with disease-induced immunity. 
The right plot in Figure \ref{p_Min_full-mod} shows the  corresponding plot when immunity comes from  vaccinating uniformly in the community (which is equivalent to disease-induced immunity for a model assuming a completely homogeneous community). In Figure S1 of the SM we show a similar plot for the model allowing for heterogeneities with respect to age and social activity but not with respect to susceptibility (treated in \cite{BBT}). The $p_{Min}$-values for its disease-induced immunity are very similar to those of the present model (left plot of Figure \ref{p_Min_full-mod}).  

In the left plot is seen that, for a fixed value of $R_0$, the necessary amount of preventive measures needed to avoid a large future outbreak decreases quite rapidly with the amount of diease-induced immunity $\hat i$, and for $\hat i$ sufficiently large the color is deep blue reflecting herd immunity.

\begin{figure}[ht]
\begin{center}
\includegraphics[scale=0.6, angle=0]{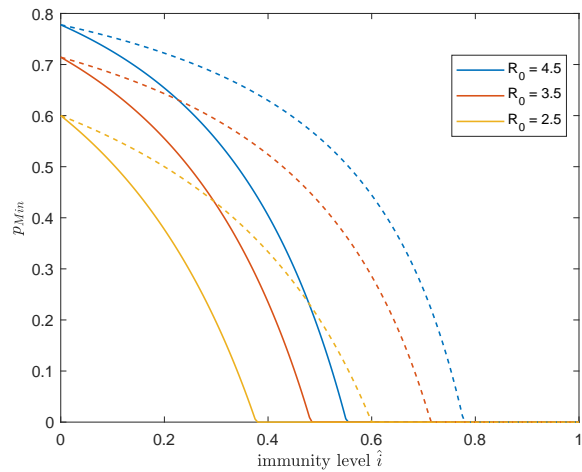}
\end{center}
\caption{Plot of the minimal amount of preventive measures, $p_{Min}$, as a function of the immunity level $\hat i$, for three different values of $R_0$. The solid curve is when $\hat i$ comes from disease exposure and the dashed curve when immunity is achieved by vaccination.}
\label{pMinplots}
\end{figure}

Figure \ref{pMinplots} illustrates the same comparison in one single figure.
For three different values of $R_0$ the minimal amount of preventive measures $p_{Min}$ is plotted as a function of the immunity level, both when immunity comes from disease exposure (solid lines) and when it is achieved by means of vaccination.

When comparing the effect of disease-induced immunity  with the effects of vaccine-induced immunity a difference is clearly observed in each of the two figures. More specifically, for a given $R_0$ and some positive immunity level $\hat i$, the necessary amount of preventive measures is substantially higher if immunity comes from vaccination as compared to disease-induced immunity (most easily seen in Figure \ref{pMinplots}). As a numerical illustration, in a region with $R_0=2.5$ that has experienced an outbreak in a mitigated situation resulting in an immunity level of $\hat i=25\%$, the required amount of preventive measures is $p_{Min} = 29\%$, whereas if instead the immunity level $\hat i=25\%$ came from (uniform) vaccination, then the necessary amount of preventive measures is $p_{Min}=1-1/(R_0*0.75)=47\%$.

An alternative way to compare the effect of disease-induced immunity with vaccine-induced immunity is to compare the doubling time between disease-induced and vaccine-induced immunity if all preventive measures are dropped at a time-point when transmission is very low. To illustrate this some further assumptions about the generation time distribution have to be made. These follow from the determinstic SEIR epidemic model and are provided in the SM. If, as above, we consider a region having $R_0=2.5$ and immunity level $\hat i=25\%$, then the doubling time for disease induced immunity equals 12.7 days whereas it equals 6.6 days if instead the immunity is vaccine-induced. (Figure S2 in the SM gives heatmaps for the doubling times as functions of $R_0$ and $\hat i$ both when immunity comes from disease exposure and when vaccine-induced.) Consequently, if all restrictions were to be lifted the epidemic would start growing quite quickly, but much quicker if immunity came from vaccination. The same qualitative result applies if restrictions are lifted only partially but still below $p_{Min}$.

\section*{Fitting to data}

We now use estimates of $R_0$ and current immunity levels $\hat i$ for a few different groups of related regions in order to compare the minimal preventive measures $p_{Min}$ of regions within each group. The regions that are compared are: Madrid vs Cataluna (containing Barcelona) in Spain, Lombardy (containing Milan) vs Lazio (containing Rome) in Italy, New York State vs Washington D.C., and the three Scandinavian capital regions Stockholm, Copenhagen and Oslo.

As mentioned earlier, the model is a simplification of the real disease spreading situation for COVID-19 by neglecting several heterogeneities (households, spatial aspects, social networks, ...) and by assuming that earlier preventive measures acted proportionally in the same way between all types of individuals. However, uncertainty in the estimated $R_0$ and $\hat i$ is believed to be much greater, so the obtained minimal preventive measures $p_{Min}$ for different regions are  to be interpreted as illustrations rather than exact numbers. Nevertheless it allows for a comparison between regions with similar $R_0$ but different immunity levels, as well as comparing regions with higher $R_0$ and immunity levels with regions having both lower.

For the sake of illustration we make the slightly unrealistic assumption that all regions have the same the same mixing between age groups, the same age-structure and the same variation in social activity and variation in susceptibility. The only differences between compared regions are the overall rate of contacts (measured by $R_0$) and the amount of preventive measures thus leading to different immunity levels $\hat i$.

As described above it is assumed that there currently is no  or low transmission. However, the results apply also to the situation where there is substantial ongoing transmission, the only difference is that then the issue is not to avoid a new wave, but to make transmission start declining.

We focus on comparing regions close to each other (making our assumption of similar mixing, age structure and variable activity and susceptibility more reasonable) and use estimates of $R_0$ and immunity levels $\hat i$ taken from the same literaure source. In the SM we explain in detail how the estimates are obtained, but in brief it is as follows. The European country-specific $R_0$ estimates are taken from \cite{FMG20publ} (except for Sweden for which the estimate is taken from its preprint \cite{FMG20}, see SM for motivation). To obtain separate estimates of the two Spanish regions we use \cite{CNE}  estimating that Madrid has about 5\% higher $R_0$ than Cataluna (a conservative estimate of the difference). For Italy, Riccardo et al.\ \cite{R20} estimates more or less identical (initial) basic reproduction numbers for Lombardy and Lazio, so here we have not distinguished between the two regions. The $R_0$ estimates for New York and Washington D.C.\ are taken from \cite{UMB20}.

The immunity levels are harder to find estimates of in the literature. For this reason we have used the official number of case fatalities per 100~000 individuals in the separate regions as of October 5, 2020, and assumed that the infection fatality risk (\emph{ifr}) equals 0.5\% (slightly smaller than the estimated \emph{ifr} for China in \cite{VOD20}). By assuming that all infected individuals become immune, and assuming no prior immunity, this gives an estimated immunity level. Of course, the \emph{ifr} most likely differs substantially between regions owing to differences in age-distribution and health care. Further, the choice to set $ifr=0.5\%$ is a rough approximation, as is the assumption that all infected become immune and that there is no prior immunity. The region-specific immunity levels $\hat i$ should hence be seen as illustrations, but their order relations are most likely correct and when this holds true the qualitative comparison statements remain true.

\begin{table}[ht]
\renewcommand\thetable{1}
\caption{Estimates of $R_0$ and current disease-induced immunity levels $\hat i$ for different regions, and the corresponding estimated minimal preventive measures $p_{Min}$. For comparison, the minimal preventive measures needed to avoid a large outbreak at the start, $p_{Min}^{(start)}$, and the minimal preventive level when immunity instead is achieved by Vaccination, $p_{Min}^{(Vac)}$, are listed.}
\centering 
\begin{tabular}{|l | c | c | c | c | c | c|}
\hline             
Region & $ R_0$ & Deaths/100k & $\hat i$ (\%) & $p_{Min}$ (\%)& $p_{Min}^{(start)}$ (\%)& $p_{Min}^{(Vac)}$ (\%)
\\
\hline                
Madrid & $4.7$ & 145  &  29.0    & 58.3  & 78.7 & 70.0
\\
Cataluna & $4.5$ & 77.4  &  15.5   & 68.9  & 77.8 & 73.7
\\
[0.5ex]
\hline                  
Lombardy & $3.4$ & 168 &  33.6  & 34.7  & 70.6 & 55.7
\\
Lazio & $3.4$ &  16.2 &  3.2  & 68.6   &  70.6 & 69.6
\\
[0.5ex]
\hline                  
New York & $4.9$ & 169 &  33.8  & 54.4  & 79.6 & 69.2
\\
Washington D.C. & $2.5$ &  89.4 &  17.9  & 40.8   & 60.0 & 51.3
\\
\hline
Stockholm & $3.9$ & 102  &  20.4   &  59.7 & 74.4 & 67.8
\\
Copenhagen & $3.5$ & 20.0  &  4.0   &  69.0 & 71.4 & 70.2
\\
Oslo & $3.0$ & 11.4  &  2.3   & 65.1  & 66.7 & 65.9
\\
\hline
\end{tabular}\label{tab_p_Min}
\end{table}
In Table \ref{tab_p_Min} the estimated $R_0$ and the disease induced immunity levels $\hat i$ are given first. Then comes $p_{Min}$ computed from the model for the estimated $R_0$ and $\hat i$ and hence taken from the heatmap (Figure \ref{p_Min_full-mod}). As a comparison the initial minimal preventive level required to avoid a large outbreak at the start (when $\hat i=0$), $p_{Min}^{(start)}=1-1/R_0$, is listed, as is the minimal preventive level required to avoid an outbreak if the current immunity level $\hat i$ was obtained instead from uniform vaccination, $p_{Min}^{(Vac)}=1-1/(R_0(1-\hat i))$. (Table S1 in the SM give the corresponding doubling times if restrictions were lifted.)

It is seen that no studied region has reached herd immunity, meaning that none of the regions can lift all restrictions without risking a new large outbreak. By comparing $p_{Min}$ with $p_{Min}^{(start)}$ it is further seen that the high levels of required preventive measures at the start of the epidemic have been reduced substantially in regions having suffered from high transmission during the first epidemic wave. More specifically, $p_{Min}$ in Lazio (includes Rome) now clearly exceeds that of Lombardy (includes Milan). Cataluna (includes Barcelona) also seems to require slightly more preventive measures than the Madrid region, but the difference is small. New York still needs more preventive measures than Washington D.C., but the difference has dropped compared to the initial required minimal levels. Among the Nordic capital regions, Stockholm had the highest initial minimal preventive measures to avoid an outbreak, whereas now Copenhagen has highest minimal preventive measures followed by Oslo, but the differences are small.

If instead $p_{Min}$ values are compared with $p_{Min}^{(Vac)}$, it is seen that disease-induced immunity plays a more significant roll as compared with immunity achieved by vaccination. In particular, the regions having highest immunity levels (New York, Madrid and Lombardy) would clearly have larger  minimal preventive requirements had immunity come from vaccination.

\section*{Discussion}

The main aim of the paper has been to compare the levels of restrictions needed to avoid new major outbreaks of COVID-19 for different regions having different initial potential ($R_0$) and different current immunity levels $\hat i$. Clearly, regions with high $R_0$ that have not yet experienced much spreading need to be most careful, but perhaps more interesting is a comparison between a region with high $R_0$ having experienced much transmission, with another region having smaller $R_0$ but also having lower immunity. The main conclusion from our study is that disease-induced immunity reduces the risk for a large future outbreak substantially more than when immunity is achieved from vaccination. Smaller local outbreaks are possible irrespective of region and are not the focus of the present paper.

In the comparison of different regions it is seen that the region now requiring the highest amount of preventive measures  may have switched from a region with high $R_0$ that has experienced high transmission, to another region having smaller $R_0$ but which has experienced less transmission.

The epidemic model studied allows for individual variation owing to age, social activity and variable susceptibility. The age effect is taken from an empirical study. However, the variation owing to social activity and variable susceptibility is chosen arbitrary but the choice is believed to be less variable (with lighter tails) than reality. Many other heterogeneities are ignored (e.g.\ households, schools and work places, spatial aspects, travel and commuting, ...) but it is believed that the effect of adding such other heterogeneities is that $p_{Min}$ is shifted close to proportionally.

A greater uncertainty lies in the estimation of $R_0$ and the immunity levels $\hat i$, but this can be reduced once better data become available. The estimates of $p_{Min}$ in Table \ref{p_Min_full-mod} are hence only to be interpreted as illustrations.

A different extension could be to consider preventive measures acting differently between different types of individual. The present framework can easily be extended to this situation, the missing information is estimates of how prevention have reduced contacts differently between different pair of subgroups of individuals.

We conjecture that our two main qualitative results hold true. These are that the effect of disease-induced immunity is greater than vaccine-induced immunity, and that regions having suffered from many infections up until now, may be in a better situation with regards to future outbreaks as compared to other regions with lower $R_0$ but with no or low immunity levels.

\pagebreak

\section*{Supplementary Materials}

\subsection*{A deterministic SEIR model and the fraction of the population infected}

In this supplementary information we describe the deterministic SEIR (Susceptible, Exposed, Infectious, Removed) epidemic model in a population partitioned by age, activity level and (relative) susceptibility. The model is an extension of the one in \cite{BBT}, and the presentation below hence follows closely (and partly copies) the model presentation in the SI of \cite{BBT}. For reasons of notational convenience we label the types (the  combination of age, activity level and susceptibility level) from $1$ to $m$, where $m$ is the product of the number of age cohorts, the number of activity levels and the number of susceptibility levels. In our model $m= 6 \times 3 \times 3 =54$. A more detailed exposition than the one presented here can be found in \cite[Sections 5.5 and 6.2]{Ande00}.

We assume that for all $j \in \{1,\cdots,m\}$ the population consists of $n_j$ people of type $j$.  We set $n = \sum_{j=1}^m n_j$ and $\pi_j = n_j/n$. We assume that the population is large and  closed, in the sense that we do not consider births, deaths (other than possibly the deaths caused by the infectious disease) and migration. Throughout the epidemic, $n_i$ is fixed, so people who die from the infectious disease are still considered part of the population.  For $j,k \in \{1,\cdots,m\}$, every given person of type $j$ makes ``infectious contacts" with every given person of type $k$ independently at rate $\alpha a_{jk}/n$. If at the time of such a contact the type-$j$ person is infectious and the type-$k$ person is susceptible then the latter becomes latently infected (Exposed). People of the same type may infect each other, so $a_{jj}$ may be strictly positive  (and often is!). Because the definition of an infectious contact includes that the contact leads to transmission of the disease, it is not necessarily the case that $a_{jk}$ is equal to $a_{kj}$. For the same reason it is possible that the relative susceptibility exceeds 1. The parameter $\alpha$ is a scaling parameter, used to quantify the impact of control measures in the main paper, without measures $\alpha$ is set so that $R_0$ has the desired value. Exposed individuals become Infectious at constant rate $\sigma$ and infectious individuals recover or die (are Removed) at constant rate $\mu$. The rates of becoming infectious and removal are assumed to be independent of type. It is straightforward to extend the model to make those rates  age and/or activity level and/or susceptibility level dependent. However,  dependence on these factors would have impact on the relationship between $p_{Min}$ and the doubling time after the first wave.

In the described multi-type SEIR model, the expected number of people of type $k$  that are infected by an infected person of type $j$ during the early stages of the epidemic is $n_k \times (\alpha a_{jk}/n) \times (1/\mu) = \pi_k \alpha a_{jk}/\mu$, where $1/\mu$ is the expected duration of an infectious period. The next-generation matrix $M$ has (for $j,k \in \{1,\cdots,m\}$) as element in the $j$-th row and $k$-th column the quantity  $\pi_k \alpha a_{jk}/\mu$.
Suppose that the next-generation $M$ is irreducible, i.e.~that for any $j,k \in \{1,\cdots,m\}$ it is possible for the infection of a type-$j$ individual to lead to the infection of a type-$k$ individual, either directly or through a chain of infectives involving other types.  It is easily seen that this condition is satisfied in our model.
The basic reproduction number $R_0$ is then defined as the largest eigenvalue of $M$; it is necessarily real and positive. If $R_0>1$, then a large outbreak is possible with strictly positive probability, while if $R_0 \leq 1$ an outbreak stays small with probability 1.

In the model under consideration, where the rates $\sigma$ and $\mu$ are independent of the type of a person, there exists a Malthusian parameter $\rho$ such that the number of infectious individuals grows initially as $e^{\rho t}$, where $\rho <0$ if $R_0<1$ and $\rho>0$ if $R_0>1$. In \cite{TBD+16} it is shown that the relationship between $\rho$ and $R_0$ is (under the assumed conditions) the same for a multitype population as it is for a homogeneous  population, where $\rho$ satisfies
$$1 = \int_0^{\infty} e^{-\rho t} \beta (t) dt,$$
with $\beta(t)$ being the expected  rate of new infections caused by a person $t$ time units after he or she was infected (e.g.\ p 12 in \cite{BP2019}). (Note that $R_0^{-1}\beta(t)$ gives the probability density function of
generation-time of the epidemic.)
In our model $$\beta(t) = R_0 \int_0^t \sigma e^{-\sigma s} \mu e^{-\mu(t-s)} ds =
\begin{cases}
	      R_0 \mu \frac{\sigma}{\sigma-\mu}(e^{-\mu t}- e^{-\sigma t})& \text{ if } \mu \neq \sigma, \\
	      R_0 \mu^2 t e^{-\mu t}& \text{ if } \mu=\sigma.
\end{cases}
$$
The growth rate $\rho$ is then the unique solution in $(-\min(\mu, \sigma), \infty)$ of
$$1=R_0 \frac{\mu}{\rho + \mu} \frac{\sigma}{\rho + \sigma}.   $$
The doubling time is given by $[\ln 2]/\rho$.

We set $S_j(t)$ to be the number of people of type $j$ that are susceptible to the disease at time $t$, $E_j(t)$ the number of people of type $j$ that are latently infected, $I_j(t)$ the number of infectious people of type $j$ and $R_j(t)$ the number of removed people of type $j$ ($j \in \{1,\cdots,m\}$). Note that $S_j(t)+E_j(t)+I_j(t)+R_j(t) = n_j = \pi_j n$ for all $t\geq 0$, because the population is closed.
Again for $j \in \{1,\cdots,m\}$, we define $s_j(t)=S_j(t)/n_j$, $e_j(t)=E_j(t)/n_j$, $i_j(t)=I_j(t)/n_j$ and $r_j(t)=R_j(t)/n_j$.

Theory on Markov processes  \cite[Chapter 11]{Ethi09} (see also  \cite[Section 5.5]{Ande00} for the single type counterpart) gives that for large $n$ the above model can be described well by the following system of differential equations (again for $j \in \{1,\cdots,m\}$):
\begin{displaymath}
\begin{array}{rllll}
\dot{s}_j(t) & = & - \frac{1}{n_j} \displaystyle\sum_{k=1}^m \alpha \frac{a_{kj}}{n} S_j(t) I_k(t) &= &
-\displaystyle\sum_{k=1}^m \alpha \pi_k a_{kj} s_j(t) i_k(t),\\
\dot{e}_j(t) & = & \frac{1}{n_j}\left( \displaystyle\sum_{k=1}^m \alpha \frac{a_{kj}}{n} S_j(t) I_k(t) - \sigma  E_j(t) \right)&= &
\displaystyle\sum_{k=1}^m \alpha \pi_k a_{kj} s_j(t) i_k(t) -\sigma e_j(t),\\
\dot{i}_j(t) & = &  \frac{1}{n_j}\left( \sigma  E_j(t) - \mu  I_j(t) \right) &= &
\sigma e_j(t)-\mu i_j(t),\\
\dot{r}_j(t) & = &  \frac{1}{n_j} \mu I_j(t)  &= & \mu i_j(t).
\end{array}
\end{displaymath}
To be complete, in the main text, we use when analysing the time-dependent behaviour of an epidemic that for all $j \in \{1,\cdots,m\}$, $s_{j}(0)=1-\epsilon$, $e_{j}(0)=\epsilon$ and $i_{j}(0)=r_{j}(0)=0$.
In the analysis below we do not impose specific assumptions on the initial conditions.

The epidemic will ultimately go extinct, because the population is closed, so for all $j \in \{1,\cdots,m\}$ we have that  $e_j(t) \to 0$ and $i_j(t) \to 0$ as  $t \to  \infty$.  Thus $s_j(t) + r_j(t) \to 1$  as $t \to \infty$. Furthermore $s_j(t)$ is non-increasing, so $s_j(\infty) = \lim_{t \to \infty} s_j(t)$ exists.

It can be shown in the spirit of \cite[Equation (6.2)]{Ande00} that for $j \in \{1,\cdots,m\}$,
\begin{equation}
\label{finalsize}
\frac{s_j(\infty)}{s_j(0)} = \exp\left[-\alpha \sum_{k=1}^m a_{kj} \pi_k \left(1-r_k(0)-s_k(\infty)\right)/\mu\right].
\end{equation}
To understand this identity  we observe first that $\frac{s_j(\infty)}{s_j(0)}$ is the fraction of initially susceptible people of type $j$ who escape the epidemic,
while the sum in the right-hand side can be written as
$$\sum_{k=1}^m n \pi_k \left(1-r_k(0)-s_k(\infty)\right) \times \alpha a_{kj}/n \times \frac{1}{\mu}
=  \sum_{k=1}^m (n_k-R_k(0)-S_k(\infty)) \times \alpha a_{kj}/n \times \frac{1}{\mu}.
$$
In words the summands read as the number of people of type $k$ that were infectious at some moment during the epidemic, times the rate at which a type-$k$ person makes infectious contacts with someone of type $j$,  times the expected time an infected person is infectious. In other words, the right-hand side is the cumulative force of infection during the entire epidemic acting on a person of type $j$. Standard theory on epidemics gives that minus the natural logarithm of the probability that a given initially susceptible person of type $j$ avoids infection is the cumulative force of infection acting on that person.
Thus \eqref{finalsize} gives that the fraction of initially susceptible people that are ultimately still susceptible is equal to the probability that a given initially susceptible person avoids infection. This argument is independent of the Markov SEIR structure of our model, and it is straightforward to generalize the results of the paper to epidemic models in which infected people have a general random infectivity profile as long as the expected shape of the infectivity profile (i.e.\ they have the same density for the generation time) does not depend on the type of the infectious person.  In particular, note that the calculation of $p_{Min}$
described below holds for this more general model.

If $R_0>1$ and the epidemic is initiated by few infectives in a large population then, conditional upon a large outbreak
occurring, the final fractions of initially susceptible people of the different types are given by the unique solution of~\eqref{finalsize}, with $s_j(0)=1$ and $r_j(0)=0$ for all $j \in \{1,\cdots,m\}$, that satisfies $s_j(\infty)<1$ for all $j \in \{1,\cdots,m\}$.

\subsection*{The population matrix}

For the age structured population and contact intensities between different age groups we used \cite{WTK06} (just as was done in \cite{BBT}). The age groups are 0-5, 6-12, 13-19, 20-39, 40-59 and 60+.   The contact matrix $A^{\dagger}$, i.e.\ the matrix with elements $\{a^{\dagger}_{jk};j,k \in\{1,\cdots,6\}\}$ is deduced from Table 1 of \cite{WTK06}. Note that the numbers reported in Table 1 of \cite{WTK06} are the expected number of contacts from a person of type $j$ with people of type $k$: $c_{jk} = n_k a^{\dagger}_{jk}/n = \pi^{\dagger}_k a^{\dagger}_{jk}$. (We use $a^{\dagger}_{jk}$ and $\pi^{\dagger}$ instead of $a_{jk}$ and $\pi$ because we already use $a$ and $\pi$ to denote the fractions in the population of the different types which are characterised by age cohort, activity level and susceptibility level, while $a^{\dagger}$ and $\pi^{\dagger}$ denote contact rates and fractions in the population of the different age cohorts only). We then  divide the elements of Table 1 by the corresponding $\pi^{\dagger}_k$ to obtain the matrix $A^{\dagger}$. (The values of $\pi^{\dagger}_j$, $j \in \{1,\cdots,6\}$, are obtained using
Appendix Table 1 of \cite{WTK06} reflecting the Dutch population in 2006; $\pi^{\dagger}_1=0.0725$, $\pi^{\dagger}_2=0.0866$, $\pi^{\dagger}_3=0.1124$, $\pi^{\dagger}_4=0.3323$, $\pi^{\dagger}_5=0.2267$, $\pi^{\dagger}_6=0.1695$.) We further multiply this matrix by a constant $\alpha$  to give us freedom to set a desired value for $R_0$.
The contact matrix  is then
$$
A^{\dagger}= \begin{pmatrix}
169.0848  &  31.4167 &  17.7946 &  34.4838 &  15.8380 &  11.4441\\
   31.4448 &  274.5499 &  32.2971 &  34.8449  & 20.6027 &  11.5031\\
   17.7911  & 32.3408 & 224.2115  & 50.7628 &  37.4995 &  14.9835\\
   34.4790 &  34.8818  & 50.7145 &  75.6476 &  49.4552 &  25.0708\\
   15.8603 &  20.5595 &  37.5465 &  49.4388 &  61.2786  & 32.9754\\
   11.4470 &  11.5503 &  14.9474 &  25.0955 &  32.9996 &  54.2119
\end{pmatrix}.
$$
As explained in the main text we can use this matrix to generate the 54 by 54 contact matrix for the model in which we take age, activity level and susceptibility level into account in the following way.
For reasons of clarity we denote the type of a person now by a three-dimensional vector $(c,a,s)$, where the first entry stands for the age cohort, which takes a value in $\{1,\cdots,6\}$, the second entry stands for social activity level, which can take values $\{1/2,1,2\}$ depending on whether the level is low, medium or high and the third entry stands for the level of susceptibility which in our example also takes values $\{1/2,1,2\}$ depending on whether the relative susceptibility is low, medium or high.

The expected number of type-$(c',a',s')$ people infected by a given infected person of type $(c,a,s)$ is then $C_{c,c'}  \times a \times a' \times s'$, where $C_{c,c'} = \alpha A^{\dagger}_{c,c'} \pi_{c',a',s'}$, where $\pi_{c',a',s'}$ is the fraction of the population with type $(c',a',s')$.

\subsection*{Calculation of $p_{Min}$}

Let $\bar{R}_0$ be the desired value of $R_0$ in the absence of preventive measures and $\alpha_0$ be the corresponding value of $\alpha$.  Then $\alpha_0$ is such that largest eigenvalue of the matrix $[\alpha_0 \mu^{-1}a_{jk}\pi_k]$ (i.e.~the matrix having element $\alpha_0\mu^{-1}a_{jk}\pi_k$ in its $j$-th row and $k$-th column) is $\bar{R}_0$.  For $\alpha \in (\bar{R}_0^{-1}\alpha_0,\alpha_0]$, let $\btau(\alpha)=(\tau_1(\alpha), \cdots, \tau_m(\alpha))$, where $\tau_j(\alpha)=1-s_j(\infty,\alpha)$ and $s_j(\infty,\alpha)$ $(j \in \{1,\cdots,m\})$ is the solution of~\eqref{finalsize}, when $s_j(0)=1$ and $r_j(0)=0$ for all $j \in \{1,\cdots,m\}$, that corresponds to a large epidemic.

Let $\alpha_*$ be such that the largest eigenvalue of the matrix $M_*(\alpha_*)=[\alpha_* \mu^{-1}a_{jk}\pi_k(1-\tau_k(\alpha_*))]$ is one and $h_D=\sum_{j=1}^\infty \pi_j \tau_j(\alpha_*)$.
Note that $M_*(\alpha_*)$ is the next-generation matrix for an epidemic among the remaining susceptible
population when the epidemic with $\alpha=\alpha_*$ has finished and all preventive measures are lifted (so $\alpha$ is then set to $\alpha_0$),
whence $h_D$ is the disease-induced herd immunity level for the epidemic with $R_0=\bar{R}_0$.  (The notation $\alpha_*$ and $h_D$ are as in \cite{BBT}.)

For fixed $\bar{R}_0>1$ and immunity level $\hat{i} \in (0, h_D)$, we obtain $p_{Min}$ as follows.
Let $\alpha_1=\alpha_0/\bar{R}_0$, so the epidemic with $\alpha=\alpha_1$ has $R_0=1$.  Note that
$\tau(\alpha)=\sum_{j=1}^\infty \pi_j \tau_j(\alpha)$ is strictly increasing on $[\alpha_1, \alpha_0]$,
$\tau(\alpha_1)=0$ and $\tau(\alpha_*)=h_D$.  Thus there exists a unique $\hat{\alpha} \in (\alpha_1, \alpha_*)$ such that $\tau(\hat{\alpha})=\hat{i}$.  Let $\hat{R}_0$ be the largest eigenvalue of the
matrix $[\alpha_0 \mu^{-1}a_{jk}\pi_k (1-\tau_k(\hat{\alpha}))]$.  Then $\hat{R}_0$ is the basic reproduction number for an epidemic with no preventive measures (i.e.~with $\alpha=\alpha_0$) among the susceptible population remaining when the epidemic with $\alpha=\hat{\alpha}$ has finished.
It follows that $p_{Min}$, the minimum amount of preventive measures to necessarily prevent a large outbreak among
this remaining susceptible population, is given by $1-\hat{R}_0^{-1}$.

\subsection*{The timing of preventive measures affect the overall fraction infected but not its composition}

The special case where the deterministic SEIR epidemic model described above has a next-generation matrix which splits up into a product with one factor depending on the type of the infector and the other factor on the type of the susceptible type is known as separable mixing \cite{DHB13}. In the separable mixing situation, two preventive measure $\{p_1(t); 0\le t\le t_1\}$ and $\{p_2(t); 0\le t\le t_2\}$ leading to the same overall fraction infected at times $t_1$ and $t_2$, respectively, also have the same fraction infected among all different types of individuals (i.e.\ the same composition of infected) at those times, as we now show.

Suppose that $a_{jk}=f_j g_k$ $(j,k \in \{1, \cdots,m\})$, where $f_j>0$ and $g_j>0$ for all $j \in \{1,\cdots,m\}$, and that $\alpha$ is time-dependent and denoted by $\alpha(t)=\alpha_0 p(t)$, where $\alpha_0$ is the value of $\alpha$ in the absence of any preventive measures.  Then the differential equation for $s_j(t)$ becomes
\begin{equation*}
\dot{s}_j(t)=-\alpha(t)\left(\sum_{k=1}^m \pi_k f_k i_k(t)\right)g_j s_j(t)\qquad (j \in \{1,\cdots,m\}),
\end{equation*}
so
\begin{equation}
\label{equ:susode1}
\dfrac{ds_j}{ds_1}=\frac{g_j s_j(t)}{g_1 s_1(t)}\qquad (j \in \{1,\cdots,m\}).
\end{equation}

Assume without loss of generality that $g_1=1$.  Then solving~\eqref{equ:susode1} yields
\begin{equation}
\label{equ:sus2}
s_j(t)=s_j(0) \left(\frac{s_1(t)}{s_1(0)}\right)^{g_j}\qquad (j \in \{1,\cdots,m\}).
\end{equation}
Consider a fixed immunity level $\hat{i}>0$ that is attainable and suppose it is achieved at time $t_0$. Then $1-\hat{i}=\sum_{j=1}^m \pi_j s_j(t_0)$ so, using~\eqref{equ:sus2}, $s_1(t_0)$ is given by the unique solution in $[0, s_1(0))$ of
\begin{equation}
\label{equ:sus3}
\sum_{j=1}^m \pi_js_j(0)\left(\frac{s_1(t_0)}{s_1(0)}\right)^{g_j}=1-\hat{i},
\end{equation}
and then $s_j(t_0)$ $(j \in \{2,\cdots,m\})$ are uniquely determined by~\eqref{equ:sus2}.  Note that if the immunity level $\hat{i}$ is fixed then the solution of~\eqref{equ:sus3} is independent of the preventive measure of $\{p(t); t \ge 0\}$ and hence so are the fractions infected among the different types when the immunity level $\hat{i}$ is reached.

The model considered in the present paper is not of the separable mixing form. More precisely, the variable social activity and the variable susceptibility enter the expression in the form of separable mixing, but the age-structure does not. However, the effect of this rather small deviation from separable mixing is negligible as Figure S3 shows. In Figure S3 we have plotted $p_{Min}$ as a function of the disease-induced immunity $\hat i$ (assuming $R_0=2.5$) for two very distinct types of preventive measures. The blue curve corresponds to a constant preventive measure $p(t)=p$ until the epidemic stops, where the value $p$ is induced by the final overall fraction infected $\hat i$. The red curve is instead obtained by having no restrictions until the level $\hat i$ is reached and at this time the epidemic is stopped (we can think of a complete lockdown). We hence have two very different preventive measures, one having constant restrictions from the start and the other strategy having no preventive measures until a sudden stop. Nevertheless we see that the two curves in Figure S3 are indistinguishable.

The consequence is that the minimal preventive measures required to prevent future outbreaks, $p_{Min}$ under a mitigated epidemic outbreak leading to an overall immunity level $\hat i$ is for all intents and purposes independent of how the preventive measures have varied over time.

\subsection*{Adding variable infectivity has no effect}

We now argue why adding variable infectivity to the other heterogeneities has no effect on the presented results as long as this variable infectivity is independent of age, social activity and susceptibility. In fact, the rate of infection from individuals of type $(c,a,s)$ acting on susceptibles of type $(c',a',s')$ depends only on the \emph{mean} infectivity of infectives of type $(c,a,s)$.  Further, when considering a deterministic model, corresponding to an infinite population, this mean is deterministic, implying that only the mean infectivity enters the equations.  The same conclusion holds in a corresponding stochastic model in the limit as the total population size $n \to \infty$, which can be shown either by appealing to the law of large numbers for density dependent population processes (see e.g.~\cite[Chapter 11, Theorem 11.2.2]{Ethi09} or \cite[Theorem 2.2.7]{BP2019} for a version with time-dependent transition rates) or in a more general setting by using the Sellke construction of the epidemic (see
e.g.~\cite[Section 6.1]{Ande00}).  Note that the model with separable mixing considered above includes the case of variable infectivity, which enters via $f_j$ $(j \in \{1, \cdots,m\})$. However, for given immunity level $\hat{i}$, the fractions of the different types infected $s_j(t_0)$ $(j \in \{1,\cdots,m\})$ when an overall fraction $\hat{i}$ of the population is
infected do not depend on $f_j$ $(j \in \{1, \cdots,m\})$ and hence are independent of any  variability in infectivity.

\subsection*{The individual variation of social activity and susceptibility}

The individual heterogeneity of social activity and susceptibility was modelled identically and assumed to be independent, independent also of the age-cohort. It was assumed that 50\% have medium or normal value, 25\% has half the value and 25\% had double the value. This choice is of course very arbitrary and made mainly in order not to have too many different types of individuals (even so there are 54 types!).

Beside the specific form, a relevant question is if the model exaggerates individual heterogeneity. We think this is not the case. For one thing, the distributions of social activity and susceptibility in the model have no heavy right tails, nor do they have values close to 0, and such tails typically have major implications (cf.\ epidemics on networks where heavy tail degree distributions alter $R_0$ dramatically \cite{PSV}).

The amount of variation is sometimes quantified by the coefficient of variation $c.v.$, defined for a positive random variable by $c.v.=\sqrt{Var(X)}/E(X)$. Our model, having values 0.5, 1 and 2 with respective probabilities 0.25, 0.5 and 0.25, hence  has $c.v.=0.48$ (note that $c.v.$ is independent of the actual values 0.5, 1 and 2 -- multiplying these values by any positive constant $k$ results in the same $c.v.$). A $c.v.$ of the order 0.5 by no means reflects an unusually high individual variation. In fact, Gomes et al. \cite{G20}, and references therein, report $c.v.$ values  between 2 and 4 (for a combination of variable susceptibility and social activity) for different diseases including COVID-19.

\subsection*{Description of $R_0$ and $\hat i$ values in Tables 1 and S1}

As explained in the main text, except for Sweden, the country specific $R_0$ estimates of the European countries given in Table 1 are taken from \cite{FMG20publ}. The $R_0$ estimate for Sweden (3.9) is taken instead from the corresponding preprint \cite{FMG20}. The reason for making this change is that in the preprint Sweden clearly had the highest $R_0$ among the Scandinavian countries, whereas in the published version (\cite{FMG20publ}) Sweden's $R_0$ estimate had dropped dramatically from 3.9 to 2.7 while Norway's and Denmark's estimates were almost unchanged (in fact both increased by 0.2). The reason for this change comes from problems with fitting the effects of Sweden's unusual preventive measures later in the epidemic and is an artefact of jointly estimating $R_t$ later in the epidemic (personal communication with Neil Ferguson).

The estimates of $R_0$ for New York (the state) and Washington D.C. were taken from \cite{UMB20}. As for the Spanish and Italian subregions, we have rescaled the country estimates from \cite{FMG20publ} by the relation between region-specific estimates of Madrid and Cataluna, and between Lombardia and Lazio. For Spain, Madrid was estimated to have about a 5\% higher $R_0$ than Cataluna \cite{CNE}, so Madrid was scaled up by 2.5\% and Cataluna scaled down by 2.5\%. For Italy the two region-specific estimates differed by only 1\%, so these estimates were left unchanged.

The number of case fatalities per 100~000 individual reported in Table 1 were all downloaded from official websites. The regions in Spain, Italy, US and Sweden were downloaded from Wikipedia October 5, 2020.  The number of case fatalities per 100k for Denmark is from Danish Infectious Disease Institut \cite{SSI} per September 14, 2020, and finally from Oslo it comes from \cite{NFHI} with fatalities up to September 23 2020. By assuming an infection fatality rate ($ifr$) of 0.5\% in all regions considered (slightly smaller than estimated for COVID-19 in China \cite{VOD20}), and assuming that all infected individuals become immune but assuming no prior immunity, this gives the region-specific immunity estimates $\hat i$ of Table 1. We again stress that these are just illustrations since several assumptions are not met. For example, the $ifr$ is most likely higher in New York (and possible also in Lombardy) as compared to the other regions. Moreover, the true $ifr$ may very well be as large as 1\% which would reduce all $\hat i$ by 50\%, though this would not change their relative sizes.

\newpage

\subsection*{Fig. S1}
\begin{figure}[ht!]
\renewcommand{\thefigure}{S\arabic{figure}}
    \begin{center}
  	\includegraphics[scale=0.8, angle=0]{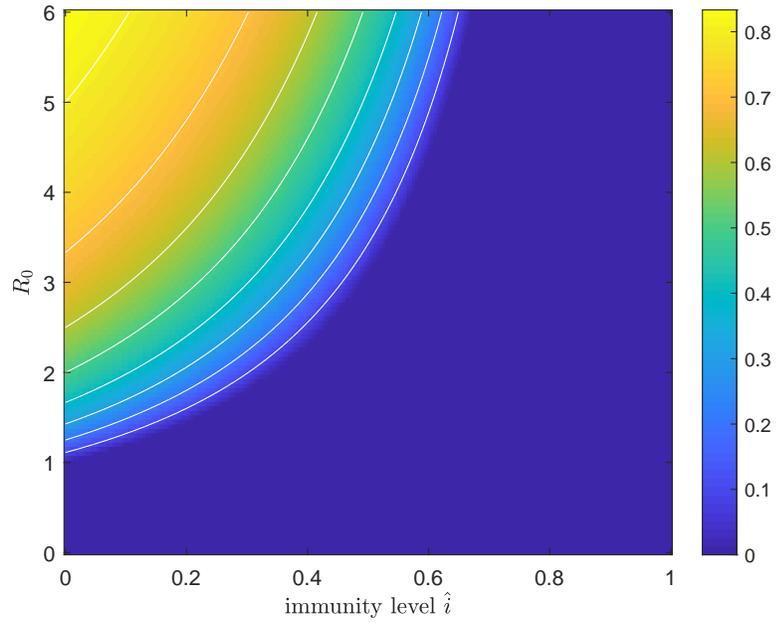}
  	\end{center}
   \caption{Plot of the minimal amount of preventive measures $p_{Min}$ for the model in \cite{BBT} allowing for heterogeneity with respect to age and social activity but not susceptibility. }
\label{heatmapscience1}
\end{figure}

\newpage

\subsection*{Fig. S2}
\begin{figure}[ht!]
\renewcommand{\thefigure}{S\arabic{figure}}
    \begin{center}
  	\includegraphics[scale=0.5, angle=0]{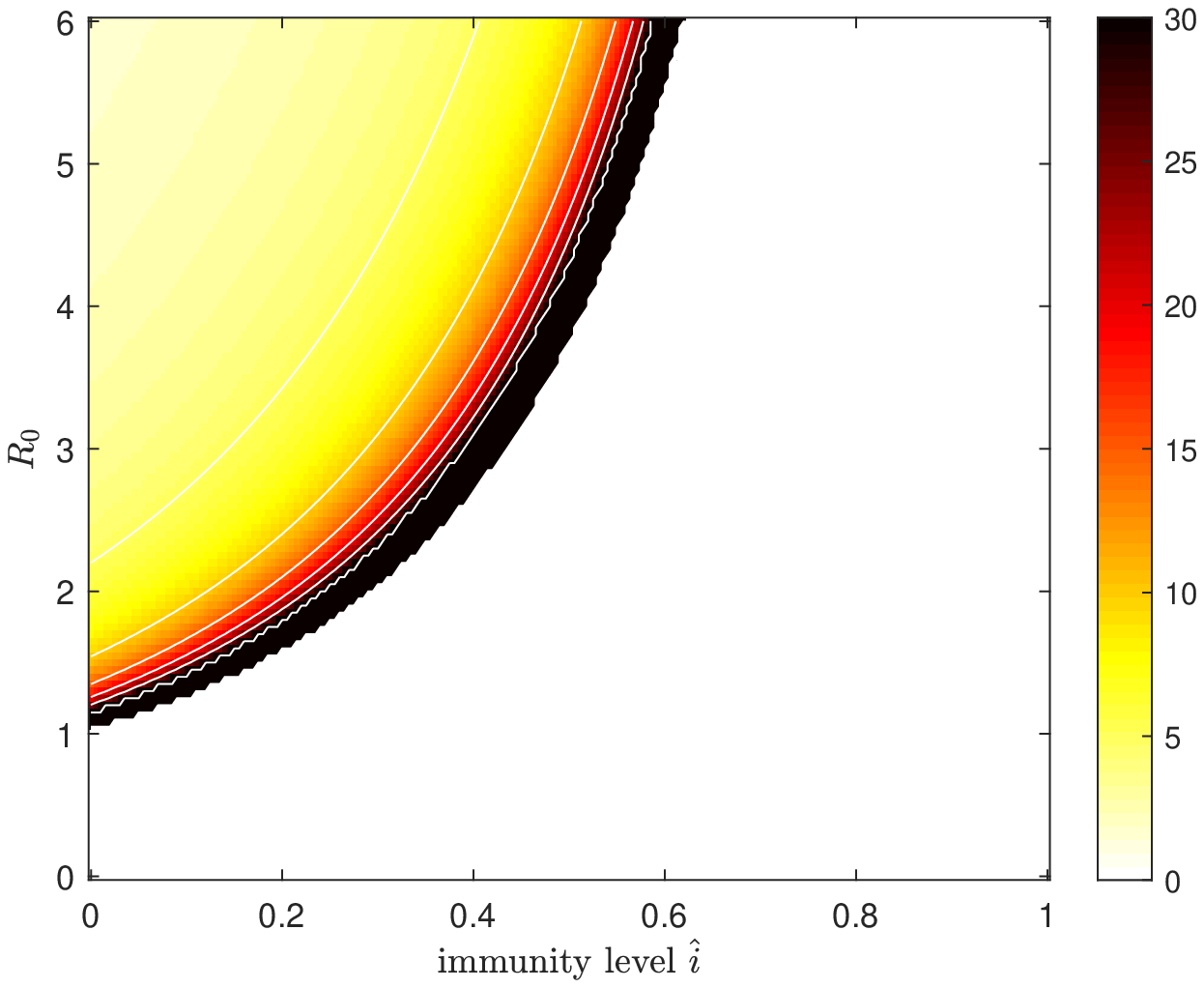}
  	\includegraphics[scale=0.5, angle=0]{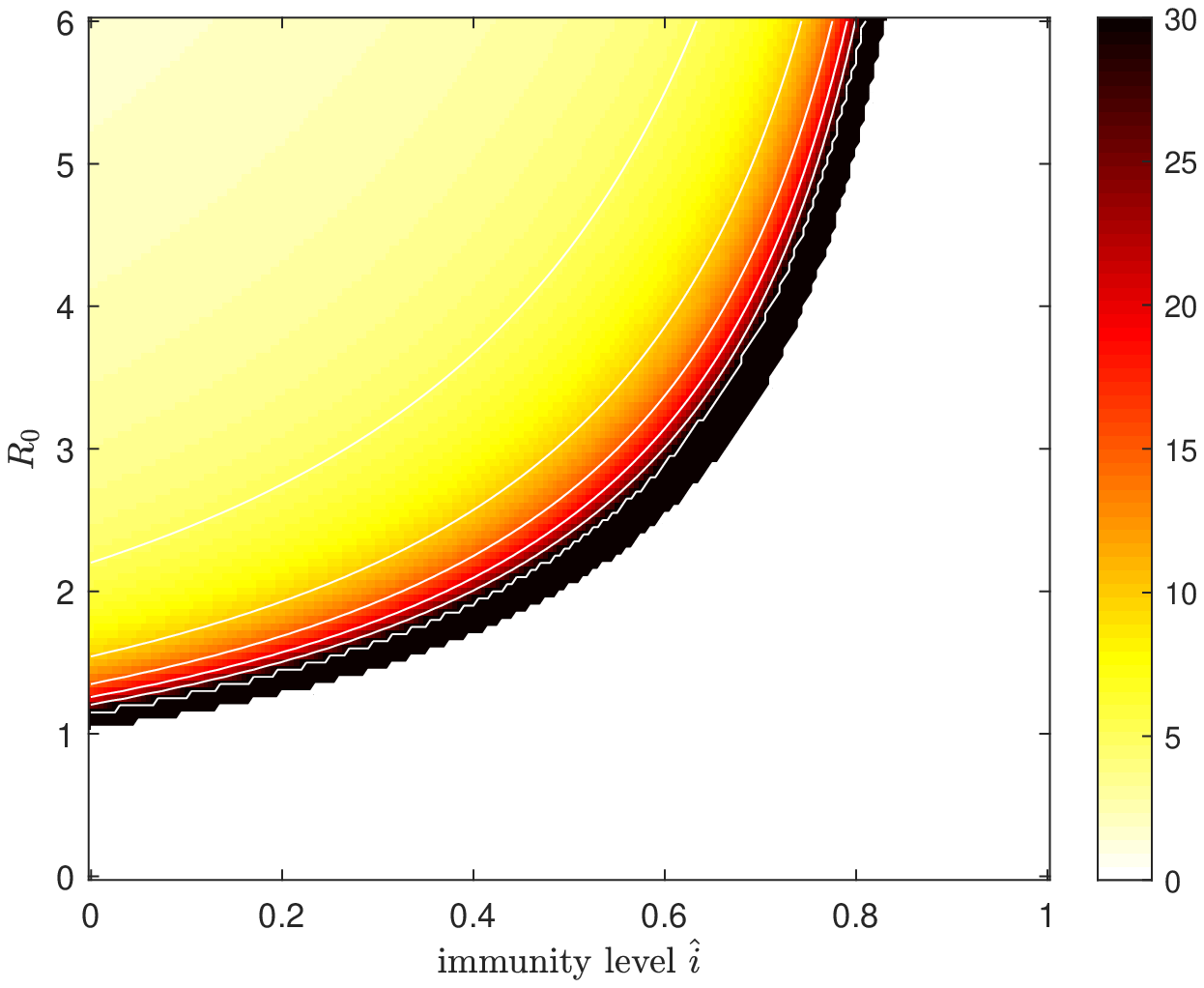}
  	\end{center}
   \caption{Heatmap for the doubling time of the epidemic growth as a function of $R_0$ and immunity $\hat i$ if all restrictions are lifted (assumptions about generation time is explained on p3 of the SM). The left plot is when immunity is disease-induced and the right when immunity is vaccine-induced. The white region is where herd-immunity is achieved meaning that an epidemic will not grow implying that there is no doubling time.}
\label{heatmapdoubling}
\end{figure}

\newpage

\subsection*{Fig. S3}
\begin{figure}[ht!]
\renewcommand{\thefigure}{S\arabic{figure}}
    \begin{center}
  	\includegraphics[scale=0.8, angle=0]{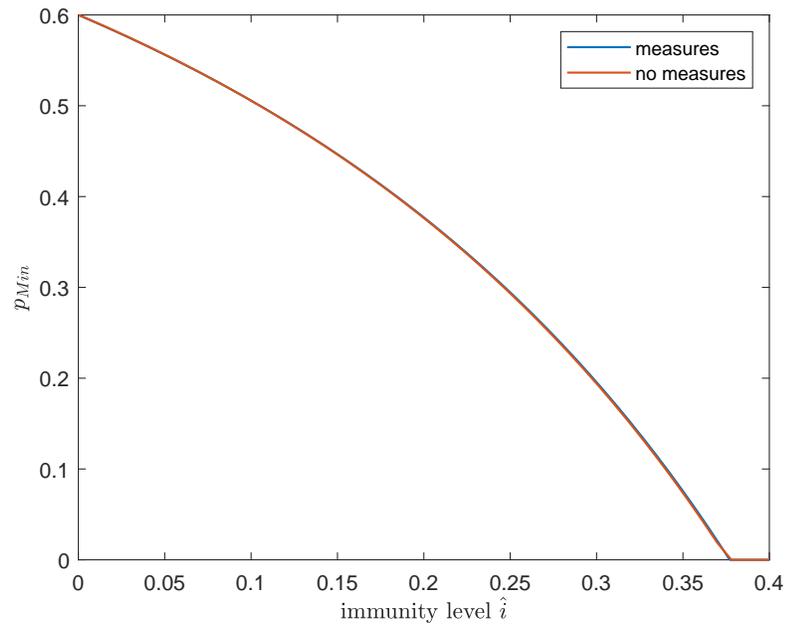}
  	\end{center}
   \caption{Plot of the $p_{Min}$ as a function of $\hat i$ assuming $R_0=2.5$. The blue curve is for the situation that preventive measures are kept constant at a level such that the outbreak ceases when a fraction exactly $\hat i$ of the population has been infected, and the red curve is obtained when there are no preventive measures until the level $\hat i$ is reached -- then the epidemic is stopped. The two curves are indistinguishable.}
\label{timevarying}
\end{figure}

\newpage

\subsection*{Table S1}

\begin{table}[ht]
\renewcommand\thetable{S1}
\caption{Estimates of $R_0$ and current disease-induced immunity levels $\hat i$ for different regions, and the corresponding doubling times $t_D$ (in days) during the exponential growth in case all restrictions were lifted at a time-point when transmission is very low. For comparison, the initial doubling times if no restrictions were put in place initially, $t_{D}^{(start)}$, and the doubling time when immunity instead is achieved by Vaccination, $t_{D}^{(Vac)}$, are listed.}
\centering 
\begin{tabular}{|l | c | c | c | c | c | c|}
\hline             
Region & $ R_0$ & Deaths/100k & $\hat i$ (\%) & $t_{D}$ & $t_{D}^{(start)}$ & $t_{D}^{(Vac)}$
\\
\hline                
Madrid & $4.7$ & 145  &  29.0    & 4.4  & 2.1 & 2.9
\\
Cataluna & $4.5$ & 77.4  &  15.5   & 3.0  & 2.1 & 2.5
\\
[0.5ex]
\hline                  
Lombardy & $3.4$ & 168 &  33.6  & 10.2  & 2.9 & 4.8
\\
Lazio & $3.4$ &  16.2 &  3.2  & 3.1   &  2.9 & 3.0
\\
[0.5ex]
\hline                  
New York & $4.9$ & 169 &  33.8  & 5.0  &  2.0  &  3.0
\\
Washington D.C. & $2.5$ &  89.4 &  17.9  & 8.1  &    4.2  &    5.6
\\
\hline
Stockholm & $3.9$ & 102  &  20.4   &  4.2 &    2.5  &    3.2
\\
Copenhagen & $3.5$ & 20.0  &  4.0   &  3.0 &  2.8  &  2.9
\\
Oslo & $3.0$ & 11.4  &  2.3   & 3.5  &  3.3  &  3.4
\\
\hline
\end{tabular}\label{tab_doubling}
\end{table}

\end{document}